This is the author's version of the work. The definite version was published in: Bajwa, S.S., Wang, X., Nguyen Duc, A. et al. Empir Software Eng (2017) 22: 2373.https://doi.org/10.1007/s10664-016-9458-0

Title of the article:

**Failures to be celebrated: an analysis of major pivots of software startups**

Authors:

**Sohaib Shahid Bajwa, Xiaofeng Wang, Anh Nguyen Duc and Pekka Abrahamsson**

Notes:

- This is the author's version of the work.
- The definite version was published in: Bajwa, S.S., Wang, X., Nguyen Duc, A. et al. Empir Software Eng (2017) 22: 2373.
- Copyright owner's version can be accessed at https://link.springer.com/article/10.1007/s10664-016-9458-0

Interested in academic Software Startup Research? Get in touch with the Software Startup Research Network, SSRN for more information, https://softwarestartups.org/

# Failures to be celebrated: an analysis of major pivots of software startups

Sohaib Shahid Bajwa, Free University of Bozen Bolzano
Xiaofeng Wang, Free University of Bozen Bolzano
Anh Nguyen Duc, Norwegian University of Science and Technology
Pekka Abrahamsson, Norwegian University of Science and Technology

Abstract
In the context of software startups, project failure is embraced actively and considered crucial to obtain validated learning that can lead to pivots. A pivot is the strategic change of a business concept, product or the different elements of a business model. A better understanding is needed on different types of pivots and different factors that lead to failures and trigger pivots, for software entrepreneurial teams to make better decisions under chaotic and unpredictable environment. Due to the nascent nature of the topic, the existing research and knowledge on the pivots of software startups are very limited. In this study, we aimed at identifying the major types of pivots that software startups make during their startup processes, and highlighting the factors that fail software projects and trigger pivots. To achieve this, we conducted a case survey study based on the secondary data of the major pivots happened in 49 software startups. 10 pivot types and 14 triggering factors were identified. The findings show that customer need pivot is the most common among all pivot types. Together with customer segment pivot, they are common market related pivots. The major product related pivots are zoom-in and technology pivots. Several new pivot types were identified, including market zoom-in, complete and side project pivots. Our study also demonstrates that negative customer reaction and flawed business model are the most common factors that trigger pivots in software startups. Our study extends the research knowledge on software startup pivot types and pivot triggering factors. Meanwhile it provides practical knowledge to software startups, which they can utilize to guide their effective decisions on pivoting.

# 1 Introduction

Startups are human institutions that create innovative products or services and search for sustainable business models under extreme uncertainty (Blank 2005; Ries 2011). Software startups are startups that build software-intensive products/services. Similar to established software companies in which software development projects have a reputation for failure (Savolainen et al. 2011), projects in software startups do fail as well. The consequence of project failure for a software startup can be even more severe than that for an established software company. This is because a majority of software startups are focused on one single project at a time. One project failure could put a software startup out of business (Giardino et al. 2016).

However, interestingly, failure is treated with positive attitude in software startups, to the extent that a few companies have the practice of celebrating each failed project, such as Supercell, according to some anecdotal evidence (Kelly 2013). Why do software startups embrace and even celebrate failures? Since the environments of software startups are extremely unpredictable and even chaotic, failures are considered a crucial way (sometimes the only way) for them to obtain important learning to validate key assumptions they make about their software products and business (Eisenmann et al. 2012). In fact, the ultimate goal of these intermediate failures along the way is to avoid the final fatal failures that put startups out of business. These intermediate failures are what we focused on in our study.

The validated learning obtained through failing fast and failing often leads software startups to making the strategic change of a business concept, product, or different elements of a business model. This type of change is called pivot in the Lean Startup approach (Ries 2011). It is claimed that pivot is the most frequently occurring commonality among different successful startups (Ries 2011). Pivot is inevitable for almost all software startups to survive, grow and eventually obtain sustainable business models. Only a few startups get their business models right immediately. It is evidenced by the fact that many successful software startups did not turn out to be what they had initially started with. For instance, Flickr used to be an online multiplayer role playing game rather than a photo managing and sharing service (Nazar 2013), while Twitter was initially developed as a podcast service, not a microblogging service (Carlson 2011).

The importance of pivot for software startups deserves research attention. However, due to the nascent nature of the research on software startups, there is a scarcity of studies on pivots in software startups. Comprehensive and valid knowledge is yet to be built on what trigger software startups to pivot, how and why they make certain pivot decisions, and how they actually pivot. The study reported in this paper is one of the first attempts to fill this knowledge gap. The objective of the study is to lay the foundations for future studies on software startup pivots by providing the basic understanding of pivots in software startups. The basic understanding includes the factors that trigger software startups to pivot, and the major types of pivots that software startups make when failures happen. To this end, the research questions that guided the study are phrased as following:

RQ1: What are the factors that trigger software startups to pivot?
RQ2: What are the types of pivots software startups undertake?

To answer the research questions, we employed a systematic research process. We collected online materials as secondary data and analysed the major pivots in 49 software startups reported in these materials, including the well-known companies such as YouTube, Flickr, Pinterest and Twitter. The online materials allowed us to quickly obtain useful data on as many significant pivots in software startups as possible. Based on the analysis of the pivots in these 49 software startups, we extracted a list of factors that triggered them to pivot, and identified a set of major types of pivots they conducted. To better structure the triggering factors and pivot types, we categorized them into different groups respectively.

The rest of this paper is organized as follows: in Section 2, the background literature and related work are reviewed. Section 3 describes the research approach employed in the study. The research findings are presented in detail in Section 4, and further discussed in Section 5. The paper is summarized in Section 6, which also outlines the future research.

## 2 Background and Related Work

### 2.1 Software Startups and Lean Startup

Software startups are challenging endeavours. A systematic mapping study (Paternoster et al. 2014) reveals the most frequently reported contextual features of a software startup: general lack of resources, high reactiveness and flexibility, intense time pressure, uncertain conditions, and tackling dynamic and fast growing markets. Software startups are dealing with various difficulties constantly emerging from different directions. Some of the top challenges include developing technologically innovative software products that require novel development tools and techniques, defining minimum viable product to capture and evaluate the riskiest assumptions that might fail a business concept, and discovering an appropriate business strategy to deliver value (Giardino et al. 2015).

Inspired by the lean principles from Toyota manufacturing and production system (Womack et al. 1990), Ries (2011) presents a new approach of entrepreneurship and innovation – referred to as Lean Startup. Lean Startup focuses on the efforts that create value to customers and eliminate waste during the development phase. However, since customers are often unknown, what they could perceive as value is also unknown. Therefore, entrepreneurs should "get out of the building" to discover customers from day one (Blank 2013). Instead of emphasizing on a business plan, Lean Startup advocates to build the product iteratively and deliver to the market for earlier feedback. The core activity of any lean startup is based on the Build-Measure-Learn (BML) loops, through which a startup turns an idea into a product, measures customer response, and then learns. This can be done through developing minimum viable products (MVP). This learning is referred to as validated learning, where each hypothesis on a business model is validated, and then a decision is made on whether to pivot or persevere. Therefore, Lean Startup is also referenced as hypothesis-driven entrepreneurship (Eisenmann et al. 2012).

## 2.2 Failure in Software Startups

In Software Engineering literature, software project failures are defined in terms of cost and schedule over-runs, project cancellations, and lost opportunities for the organizations that embark on the difficult journey of software development (Linberg 1999). While software project failures can lead to business failures in established companies, it is not so drastic as in the context of startup where one failed project could put a startup company out of business (Giardino et al. 2016), the eventual failure of a startup when it passes a point of no return, leading to the termination of business. Software startup failure rate can be as high as 75 % to 90 % (Nobel 2011; Marmer et al. 2011).

The essence of Lean Startup methodology is to help startups make early and cheap failures as often as possible, and learn from these intermediate failures in order to avoid final catastrophes (Ries 2011). These intermediate failures that occur during the courses of startup processes are the ones that should be embraced actively and that can lead to pivoting, therefore the focus of our study. The related work is also reviewed based on this perspective on failure in software startups. Learning from this type of failures is crucial. However there is a paucity of studies on failures in software startups. One exception is Giardino et al. (2014), in which two software startup failures were documented and the reasons identified. One main reason is not changing directions when they were needed, or in other words, necessary pivots were not taken at due time. This issue is echoed in the study of Shepherd et al. (2009). According to them, the moment of failure is not always that straightforward. Sometimes entrepreneurs decide to continue the business even though the situation is hopeless. However, in some other cases, entrepreneurs do pivot and improve their entrepreneurial learning experience, but there is no study that investigates factors that trigger pivots in software startups, to the best knowledge of the authors.

## 2.3 Pivots in Software Startups

Pivot is often considered the synonym of change. However, it is not about introducing just any change and making any decision. Several definitions of pivot are presented in literature in recent years. Pivot is considered as validating a hypothesis related to a business model (Blank 2005; Maurya 2012), even though it is not compulsory that a pivot can only be related to a business model. According to Ries (2011), a pivot is a special kind of change designed to test and validate the assumptions about a product, business model and the engine of growth. Based on these definitions, for this study we define a pivot as a strategic decision which leads to the significant change to one or more, but not all, elements of a startup: product, entrepreneurial team, business model or engine of growth. When all of these elements change at the same time, it is not considered a pivot but starting a completely new and different business.

Previous research on pivots in software startups is limited (Paternoster et al. 2014). Bosch et al. (2013) offer an alternative to pivot or persevere i.e., to abandon the idea, by presenting a software development model for early stage software startups. However, the study is not primarily focused on pivoting. The study by Van der Van and Bosch (2013) describes pivots that software startups have made, and couples them with architecture decisions. It compares pivots and software architecture decisions in developing a new product, and presents the similarities and differences between these two types of decisions. According to the study, both pivots and software architecture decisions consider risk as a triggering factor in making a decision, while the focuses of pivots and software architecture decisions are different. This study considers a pivot an example of business decisions only, and does not consider product related pivots. Another study by Hirvikoski (2014) provides an overview of how software startups pivoted historically through the examples of Twitter, Google and Facebook. The study argues that most successful startups have made multiple pivots during their journey. However the pivot examples are not based on empirical data, and the study does not shed lights on why startups pivot. Moreover, it lacks rigorous scientific argumentation. Terho et al. (2015) identify different pivot types (product zoom-out, customer segment, business architecture etc.), and explain how they affect the different parts of the lean canvas model. However, there is a lack of information on how a pivot is identified and categorized under a specific pivot category.

There is a scarcity in the literature to identify major pivots that software startups have made. In order to ground our research on some basis, we used the pivot types reported in Ries (2011), with the awareness that these types are subject to systematic and scientific validation. Ries (2011) presents ten different types of pivots that can happen in startups:

- Zoom-in Pivot: A single feature of a product becomes the whole product, such as a chatting feature of an online game becomes a stand-alone messager app.
- Zoom-out Pivot: Opposite to zoom-in pivot, a whole product becomes a single feature of a much larger product. For example, a photo-sharing app is extended to an social media platform for photographers.
- Customer Segment Pivot: It is to shift from one customer segment to another, e.g. a training app orginally targetting at professional atheletes later on at amateurs, because a product hypothesis is partially confirmed, solving the right problem but for different customers than initially anticipated.
- Customer Need Pivot: As a result of getting to know customers extremely well, sometimes one realizes that the problem they are trying to solve is not important for the customers, but they often discover other related problems that are important for them and can be solved.
- Platform Pivot: It refers to change from an application to its supporting platform or vice versa, e.g., shifting from an online shop to a platform that hosts online shops.
- Business Architecture Pivot: In this pivot, a startup switches business architecture e.g. going for high margin, low volume instead of focusing on mass market.
- Value Capture Pivot: The methods that capture the value a company creates are commonly referred to as monetization or revenue models. A startup can capture value it creates through different ways. An example of value capture pivot can be an online service changing from freemium price model to monthly subscription fee model.
- Engine of Growth Pivot: Typical growth engines for startups are viral (through word-of-mouth), sticky (attracting users to stay with a product/service as long as possible) and paid growth models. A startup changes its growth strategy to seek rapid and more profitable growth.
- Channel Pivot: It is a recognition that a startup company has identified a way to reach their customers more effective than their previous one, e.g., from selling a product/service via post mails to selling on online shops.

- Technology Pivot: A startup delivers the same solution by using completely different technology, e.g. an app shifting from iOS to Android platform.

In addition to these pivot types, Hirvikoski (2014) proposes a new one - social pivot, "where active changes in social factors, such as persons and environments, change the direction of a company." Similarly, there is a severe lack of scientific argumentation and examples that support this new pivot type.

## 3 Research Approach

This study is designed to be exploratory due to the nascent nature of the software startup research area and the limited number of previous studies on pivots. In order to quickly obtain useful data and understand the directions of inquiry in future primary research, we decided to use secondary data on the software startup pivot examples that we could find on different websites, to develop an initial understanding of the phenomenon under the study.

Secondary data means that the research data is either collected by individuals other than the researchers who conduct the study, or for any other purposes than the one currently being considered, or often can be a combination of these two (Vartanian 2011). There can be several sources of secondary data e.g. census, magazines, newspapers, blogs, reports etc. The advantage of using secondary data is that data collection process can be fast, and inexpensive (Vartanian 2011). Used with care and diligence, secondary data can provide a cost-effective way of gaining an initial understanding of research questions. Secondary data analysis is also considered a starting point for other research methods, often helpful in designing subsequent primary research and can provide a baseline with which to compare the primary data analysis results (Boslaugh 2007). The use of secondary data is quite common in other disciplines, such as psychology (Trzesniewski et al. 2011), and it is also becoming a suitable approach in the software engineering research community (e.g. Wang et al. 2012).

The overall research methodology employed in this study can be considered the case survey method (Yin and Heald 1975; Cruzes et al. 2015), since it "works best when the studies consist of a heterogeneous collection of case studies" (in our case, a collection of software startup pivot examples), and the researchers' main task is "to aggregate the characteristics, but not necessarily the conclusions, of these cases" (Yin and Heald 1975, p. 371). The case survey method enables the researchers to note the various experiences found in each case and then to aggregate the frequency of occurrence of these experiences, therefore ensures the analysis of qualitative evidence in a reliable manner. Cruzes et al. (2015) list case survey as one of the methods for the synthesis of qualitative and mixed-methods evidence that can be applied in Software Engineering research. To further ensure the data collection and analysis process is systematic and reliable, we adopted the systematic literature review guidelines by Kitchenham (2007) and adapted them to our context. We developed a secondary data search and analysis protocol. The protocol helps to reduce the researchers' bias, because without using it, it is possible that the data collection or analysis could be driven by researcher expectations (Kitchenham 2007).

The overall data collection and analysis process employed in the study is illustrated in Fig. 1 and explained in detail in the following text.

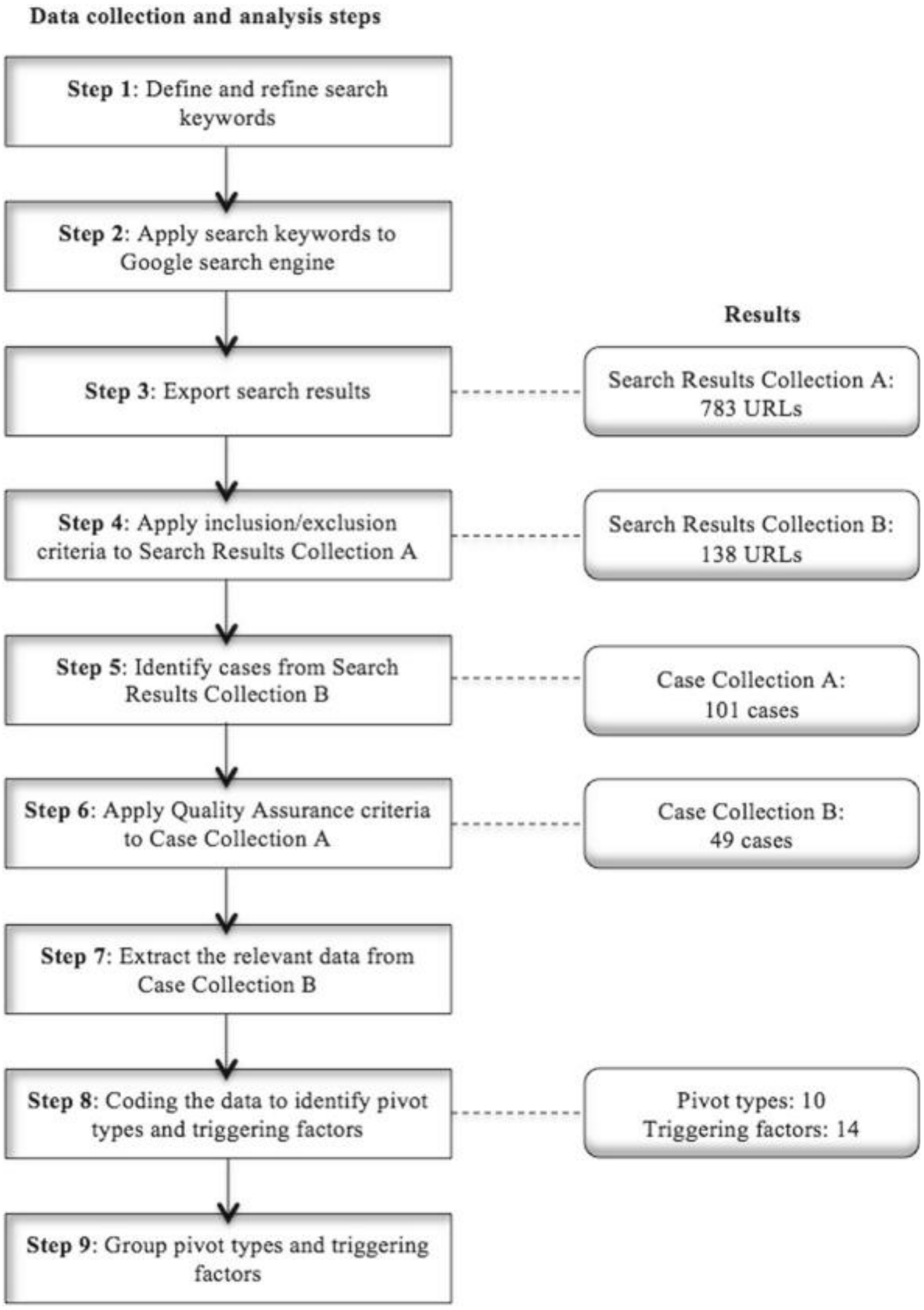

Fig. 1 The data collection and analysis process

## 3.1 Data Collection Steps

Step 1. Define and refine search keywords

The first step of the data collection was to define the search keywords used to search the secondary data. Based on the main objectives and research questions, we brainstormed the initial set of search keywords. The search string was structured using the guidelines given by Kitchenham (2007). To ensure that we captured the keywords related to software startups, we consulted the search string used in a systematic mapping study regarding software startups (Paternoster et al. 2014). We conducted several trial searches, observed the search results, and refined the search string subsequently. As a result, the following final search string was formulated:

"startup" OR "start−up" OR "early−stage firm" OR "early stage firm" OR "early−stage compan*" OR "early stage compan*" OR "venture" AND "pivot"

We used "startup" rather than "software startup" due to the fact that often software startups are described as Internet startups, tech startups or simply startups in online sources.

We are aware that we may miss some sources that represent pivots but do not use this specific term. However this risk was mitigated by the fact that, together with the Lean Startup movement, pivot became a commonly acknowledged and used term in startup communities.[1,2]

Step 2. Apply search keywords to Google search engine

To search online sources, Google search engine was used through Chrome browser. To avoid the influence of geographical location on the search results the website www.google.com was used. The search was conducted by one researcher. Before starting the search process, the researcher deleted the search history in the Chrome browser, cleared browser cache, and logged out from his personal Google account. The intention of these steps is to ensure the least possible influence of personal and historical data on the search results. In the Google search settings, BGoogle instant predictions^ was turned off, and B100 results/links per page^ was enabled.

The search was conducted on the first author's laptop on February 26th, 2016. The search resulted in 1,070,000 hits. However, Google search engine does not show more than 1,000 results per search query (Jerkovic 2010). In addition, it omits the results that it considers similar or duplicates by default. We disabled this omit option to include those results to be analysed manually later. As a result, in total 783 results were displayed and eventually accessible.

Step 3. Export search results

The search results needed to be exported in order to be analysed by multiple researchers. For this reason, the first author installed SEOQuake[3] plugin to his Chrome browser, which automatically exported the search results (in the format of URLs) into an Excel file. This is the Search Results Collection A which contains 783 URLs and each points to a webpage.

Step 4. Apply inclusion/exclusion criteria to Search Results Collection A

To select the webpages that contain relevant and reliable content for this study, we applied a set of inclusion/exclusion criteria to Search Results Collection A.

The inclusion criteria are:

- The URL is working, and freely available (or accessible)
- The topic of the webpage is about pivoting in startup context
- The webpage contains examples of startup pivots

[1] https://www.startupgrind.com/blog/is-pivot-the-new-fail/
[2] http://www.inc.com/alan-spoon/what-pivot-really-means.html
[3] http://www.seoquake.com

- The pivot examples are coming from software-based startups
- The webpage is in English

The exclusion criteria are:

- The webpage contains the duplicated content of a previously examined webpage
- The webpage is non text-based (e.g. videos, audios, or images)
- The webpage on Slideshare, Quora, LinkedIn, personal (or company) blogs

We excluded Slideshare because of the synthetic content and lack of contextual information, whereas webpages from Quora, LinkedIn and personal (or company) blogs are excluded for the potential subjectivity in the content.

To decide if a startup is software startup and if the pivot described is a real pivot, we used the definition of software startup (defined in the beginning of Section 1) and pivot (defined in Section 2.3) to guide the inclusion/exclusion step.

The first two authors conducted this step separately. The evaluation results from the two researchers were compared and the disagreed items (5 %, 39 were discussed out of 783 URLs) were discussed between the two researchers until a consensus was achieved. This step resulted in the Search Results Collection B which contains 138 URLs and represents 138 webpages.

Step 5. Identify cases from Search Results Collection B

We read through the content of the 138 webpages, and looked for the information about the software startups that pivoted during their startup processes. We considered each mentioned software startup a potential case for further analysis. Since this step was relatively objective and straightforward, it was mainly conducted by the first author. In the case of doubt, the second author was consulted. This step resulted in the Case Collection A that contains 101 cases. The 138 webpages were re-organized according to the identified cases.

Step 6. Apply quality assurance criteria to Case Collection A

To ensure that we have sufficient and adequate data on the cases for further analysis, we evaluated the quality of data we had on the 101 cases in Case Collection A based on the following quality assurance criteria:

- Does the data about a case startup allow the researchers to re-construct the pivoting story of the startup in terms of what the startup was focused on before and after a pivot, and why it made the pivot?
- Do the researchers have to make excessive guessing in order to understand the pivoting type and the factors triggering those pivots?

A case is included if the answer to the first criterion is positive and the answer to the second one is negative. The first two authors conducted this step separately. The evaluation results from the two researchers were compared and the disagreed items (36 out of 101 cases) were discussed until a consensus was achieved. This step resulted in Case Collection B which contains 49 cases that were used in the data analysis.

The data regarding these cases are contained in 47 webpages. The data on one case may be spread in more than one webpage, and one webpage may contain data on more than one case. The 47 webpages (represented by their URLs) used for analysis were documented and available at a permanent address at https://figshare.com/s/152f52bb036cc6a67526.

### 3.2 Data Analysis Steps

Step 7. Extract the relevant data from Case Collection B

For each case (software startup) contained in Case Collection B, we were looking for the following information on the case:

- Background information

– Name of the startup
– Location of the company
– Founding year and/or first product release date
– Business domain
– The main business/product/service before a pivot
– The main business/product/service after a pivot
– Description and explanation on how and why the startup pivoted

To get the background information, we first used the URL obtained through our systematic search; if no information was found, we checked a startup's homepage or LinkedIn page; if still no information was found, we resorted to Wikipedia. Wikipedia is used in six cases (Docker, Fab, Seesmic, Shopify, Site59, Voylla). If there was more than one link that discussed the same software startup (e.g. two links discussing Twitter as an example of pivoting), we included all links along with the descriptions under the same startup name.

This step was conducted by the first author alone as it was mainly concerned with data retrieval. The first and second authors discussed 3 unclear cases, to resolve uncertain aspects regarding the background information of these cases.

Step 8. Coding the data to identify pivot types and triggering factors

The data extracted on each case was analysed qualitatively to identify the pivot types and the factors that triggered the reported pivots. We relied on the explanations given in the case material to identify the triggering factors of pivots. The way we selected the cases ensured that the triggering factors that led to pivots were described.

A completely open coding process was used to allow the emergence of the triggering factors, meanwhile a seed category of pivot types as described in Section 2.3 was used in the coding process to facilitate the identification of the types of pivots these software startups have experienced. Table 1 presents an example of how coding was conducted:

This analysis step was conducted by the first and second authors separately. The coding results from the two researchers were compared and the disagreed items (the pivot types of 12 pivot instances and the pivot triggering factors of 5 pivot instances) were discussed between the two researchers until a consensus was achieved.

This step resulted in 10 pivot types and 14 triggering factors.

Table 1 A coding example to identify pivot type and triggering factor

| Software Startup | Before Pivot | After Pivot | Pivot Type (cod) | Reason of Pivot | Triggering Factor (cod) |
|---|---|---|---|---|---|
| BranchOut | "*Social networking site for professionals*" | "*Talk.co. A chatting app for employees*" | product zoom-in | "*user acquisition slowed, people were only using the site occasionally*" (retention low) | Negative customer reaction |

Columns having (cod) in the title contain the codes

Step 9. Group pivot types and triggering factors

To provide a better structure of the pivot types, we classified them drawing upon the four dimensions that are claimed to be vital for successful ventures by MacMillan et al. (1987) and employed in other related studies (e.g. Giardino et al. 2014):

- Product dimension: startups are developing technologically innovative solutions (Sutton 2003).
- Market dimension: it refers to identifying the essential need of the customers (Blank 2005).
- Financial dimension: it is related to the funding, investments, and return on investments and also the way a startup evolves sets the company growth and its place in market (Yu et al. 2012).
- Team dimension: it is the main driving force behind several entrepreneurial activities related to product and business development (Giardino et al. 2014).

The triggering factors were grouped into external and internal factors. External factors are those that are beyond the control of a startup, whereas internal factors stem from the decisions or activities of a startup itself.

# 4 Results

## 4.1 Description of the 49 Pivoted Software Startups

The 49 software startups included in our case survey come from all over the world, however the majority (37) are based in the United States, and 4 in Canada. Two case companies are located in Israel, while the other 4 are located in United Kingdom, Australia, New Zealand and India. For two companies we could not obtain the information on their geographic locations.

Social networks (30.61 %), e-commerce (24.44 %), and finance and business (12.24 %) are the main business domains these software startups come from. The other domains include digital government, operating system, health and travel industries. Most products developed by these startups are market-driven and Internet-based. The targeted customers are either general (such as Twitter, Yelp and YouTube) or from a specific segment (e.g., Ignighter targets at Indian users primarily).

Twenty four out of the 49 cases are recent software startups, either being founded or releasing their first products in the past five years (between 2010 and 2015), while 13 cases have launched their products during 2005–2009. 7 startups released their products first time during 1998 to 2004, while for 5 cases the product release dates we could not obtain information.

Table 2 lists the 49 software startups included in our study, including their company names (at the time their pivots were reported in the webpage), the main business ideas before and after their pivots. From these 49 cases we identified 55 instances of pivots, which means that some cases contain more than one pivot example. These cases are marked in Table 2.

### 4.2 Overview of Pivot Triggering Factors

Table 3 describes the major factors identified from the 49 cases that act as triggers for software startups to pivot. These triggering factors are grouped under either External or Internal, as defined in Section 3.2, Step 9. It is likely that a pivot instance is triggered by more than one factor.

### 4.3 Overview of Pivot Types

The major pivoting types identified in the 49 cases are listed in Table 4, organized under the dimensions of "product", "market" and "others". (Note that our findings did not reveal any pivot that can be classified as financial or team related pivots.) One pivot instance is classified under one pivot type only.

As shown in Table 4, market related pivots (45.40 % of the total 55 instances of pivots in 49 cases) are the most common types of pivots among the software startups included in this study. Another 31 % of the pivot cases are product related pivots. The results also reveal several new types of pivots (23.60 %) in addition to the existing pivot types presented in Section 2.3, including market zoom-in pivot, complete pivot and side project pivot.

In the following sub-sections, the 49 cases are classified according to their pivot types and corresponding triggering factors. Tables 5, 6 and 7 list the cases that contain one instance of pivot per case. Table 8 shows the cases that have multiple pivot instances. For each pivot type and triggering factor, we highlight the more illuminative and interesting cases in the text by providing more detailed description and insights on these cases. The direct quotations from the cases are also included while writing the description of each case, which are referred to in the text as [casename] or [casename:link number] if a case has multiple URLs. The case names and link numbering can be found at https://figshare.com/s/152f52bb036cc6a67526.

### 4.4 Product Related Pivots

In this and the following sub-sections, we use the pivot instances identified from the 49 cases to further illustrate each pivot type and the corresponding triggering factors. Table 5 shows a list of software startups that each contains one pivot instance only and the instance can be classified under the product related pivots. The triggering factors that caused these startups to pivot are also shown in the table.

#### 4.4.1 Zoom-In Pivot

Three startups - Flickr, Slack and Voylla - did product zoom-in pivot because their users appreciated one particular feature rather than the whole product they offered. Flickr is a representative example. It originally was an online massive multiplayer role-playing game called Game Neverending. It failed to attract the customers' attention. However, the game provided a photo sharing tool to allow players to share photos and save them on a webpage while playing. This turned out to be the most popular aspect of the game. The founders decided to leverage this popularity and pivoted towards a photo sharing application now

Table 2 The list of 49 pivoted software startups

| Software startups | Before pivot(s) | After pivot(s) |
|---|---|---|
| Android | Operating system for cameras | Operating system for smartphone (mobile handsets) |
| appMobi | Flycast: a mobile app selling creative audio/video banner ads for iPhone, Blackberry and Andriod devices | A set of tools for cross-platform development of mobile apps |
| BranchOut | Social networking site for professionals | Talk.co: a chatting app for employees |
| BraveNewTalent | Social recruitment platform | Social learning platform |
| Carbon | On demand valet parking | Valet parking offering several services (e.g. picked up laundry) called delivery concierge service for car |
| ChartBeat | Firef.ly: enabling website owners to see how users are mousing around their websites | Infographics about users visiting the websites |
| Citivox | Government enterprise software suite | Community-organizing tool where people can share what they like |
| Curios.me | CrowdRally: Facebook fan site network | Website for questions and answers targeting at college students |
| Docker | dotCloud: renting software and hardware (platform as a service) | Open source development system |
| Eden | Providing help to solve technical issues of consumers | Solving technical issues especially focusing on companies |
| Elto | Tweaky: marketplace for developers to make small changes in websites | Adding more functionalities and helping small business to grow by connecting them to marketers and growth strategists |
| Fab | A social network targeted at gay community | A daily flash sales website for modern and latest fashion clothes, housewares, accessories, clothing, and jewelry |
| Flickr | Neverending: a massive multiplayer online role-playing game | Sharing photos online |
| Groupize | A travel startup focusing on consumers | Group-booking solution for businesses (travel management companies etc.) |
| Groupon | thepoint.com: social goods fund raising site based on tipping point | Group buying site working on same tipping point |
| Handmake Me[a] | Host My Portfolio: a portfolio service for professional creatives | Reverse marketplace for handmade gifts |
| Hopper | A travel discovery app to find where to travel. | An app to suggest when is the best time to fly. |
| Ignighter | A group dating site for all | A group dating site for Indian users |
| Instagram[a] | Burbn: location based service | Photo sharing app having different filtration criteria |
| Jammber | Social app with the goal to become LinkedIn for the music industry | SaaS accounting platform for artists to manage documentation (filling tax forms etc.). |
| Keas | Personalized care plans for customers | Workplace wellness programs for different businesses |
| Life On Air | Air: video broadcasting | Meerkat: a mobile app to stream live video to their Twitter followers |
| MishGuru[a] | Creating and printing 3D horse shoes | A content management system |
| NextBigSound | A website where music lovers could create fantasy record labels to Bsign^ artists | An enterprise data and analytics company for artists, producers etc. |

Table 2 (continued)

| Software startups | Before pivot(s) | After pivot(s) |
|---|---|---|
| Nextdoor | Fanbase: wikipedia for sports | A social network based on neighborhood |
| Now in Store | An app that allows brand to automatically showcase their products in a portfolio | An app to help businesses to create marketing content |
| Paypal[a] | Confinity: helping Palm Pilot users exchange money electronically | Online monetary exchanges |
| Pinterest | Window shopping using mobile | Collection of favorite items, and sharing it with friends |
| RenentionScience[a] | A platform for independent artists to promote niche brand on social media | A platform that enables online businesses to re-engage existing customers |
| Seesmic | Video-based Twitter, enabling users to broadcast video clips | Social media client application to manage multiple accounts on different social platforms |
| Shopify | Online snowboard business | Online shopping cart for small businesses |
| Signpost | Deal site like groupon | Helping small and medium enterprises to market online |
| Site59 | Creating mini vacation packages from air travel, hotel accommodation, and other travel services, and selling them online. | Offer vacation packages by following the B2B2C (business to business to consumers) strategy |
| Slack | Online role playing game | Internal chat tool |
| Socrata | A cloud-based database for small-to-medium sized businesses | Cloud-based offerings for open data governments |
| Streamline | Streamlining customer service industry by cutting waiting time during call-ins | Tools to help retailers in their decision to expand |
| StyleZen | Shopping site | Technologies to help businesses leverage Pinterest as a marketing vehicle |
| SymphonyCommerce | A social shopping website that sells style, home, beauty and living goods | E-commerce platform |
| Tagged | A social network | A social discovery product |
| Turntable.fm | Stickybits: a mobile bar code scanning startup | A social media website allowing users to interactively share music |
| Twitter | Odeo: personal podcasting service | A microblogging platform |
| Vidyard | A startup making marketing videos | Providing analytics about videos |
| Voylla | Online retailers for women's apparel, jewelry, and accessories | Online retailers for women's jewelry and accessories, excluding apparel from initial offerings |
| Wix | Flash website builder | HTML5 based platform for website development |
| Woot | A electronics wholesale distributor | A unique model for online shopping (Internet retailers) |
| Workible[a] | HireMeUp: a job site helping job seekers to find jobs based on their availability | Mobile based solution to find quality people fast |
| Yelp | Automated system for email recommendations to friends | One-stop shop for local business reviews |
| YouTube | Video dating platform | Sharing videos online |
| Zealyst | Helping people build meaningful new connections and create stronger social networks | Organizing events where people can make new professional and personal connections |

[a] case containing more than one pivot instance

Table 3 Major factors triggering pivots in software startup

| Triggering factors | Description | # of pivot instances |
|---|---|---|
| External | | |
| Negative customer reaction | It refers to slow customer acquisition, slow customer retention, no or negative response from customers etc. | 15 |
| Unable to compete with competitor | Several competitors (e.g. big companies, other startup companies) outplay the startup by working on the same idea more effectively. | 5 |
| Technology challenge | Several challenges related to technology, including limitation with existing technologies (e.g. performance issues), better technology availability due to emergence of disruptive technologies. | 5 |
| Influence of investor/mentor/partner | Suggestion or pressure from investors, mentors or partners to change the direction. | 4 |
| User appreciation of one particular feature of the product | Users appreciate one specific feature, rather than showing interest in the whole product. | 4 |
| Unanticipated use of product by users | Users use product in an unexpected manner, which was not foreseen before. | 4 |
| Wrong timing | Providing a solution which market is not yet ready to accept. | 3 |
| Positive response from an unforeseen customer segment | Among different customer segments, one specific segment shows more interest in the product. | 3 |
| Running into legal issue | Legal problems with other companies (e.g., copyright issues). | 1 |
| Side project more successful than main project | Lack of interest from customers in the main project, but they are interested in the side project. | 1 |
| Targeted market narrowing | The initially targeted market becomes smaller for the startup to survive and grow. | 1 |
| Internal | | |
| Flawed business model | High cost of customer acquisition, or revenue model is not working. | 7 |
| Identification of a bigger customer need through solving an internal problem | While developing a solution internally, to support the core product, the startup realizes that the identified internal problem is the real pain point for the customers, compared to the problem their original product solves. | 5 |
| Unscalable business | Solving a problem in which not many people are interested, resulting in unscalable business. | 5 |

known as Flickr. In the case of Voylla which provides an e-commerce solution, the product zoom-in pivot happens at the content level (offering less online products) rather than at the underlying platform level. This is different from the software feature zoom-in pivot in pure software startups like Flickr and Slack.

BranchOut and Shopify did zoom-in pivot for different reasons. Starting as a social networking site for professionals, a so-called "LinkedIn meets Facebook" venture, BranchOut had a good start but user acquisition was slow and users were only using the site occasionally. New direction needed to be found, and it came to the attention of BranchOut that the messaging service was used by a lot of users so it decided to break this feature out (called Talk.co). Instead, Shopify narrowed their online snowboard business down to offering specific online shopping cart solution, because in 2004 when they needed an online shopping cart for their online business, they found no suitable choices, had to create their own, and realized that it was the same issue many other small companies ran into.

Table 4 Major pivot types in the software startups

| Dimension | Pivot type | # of pivot instances |
|---|---|---|
| Product | Zoom-in: a single feature of a product becomes the whole product. | 7 |
| | Technology: the same solution using comletely different technology. | 5 |
| | Platform: a product becomes a platform or vice versa. | 3 |
| | Zoom-out: a whole product becomes one feature of a much larger product. | 2 |
| Market | Customer need: switch to a different problem that customers have | 17 |
| | Customer segment: switch to a different customer segment than the one originally conceived. | 6 |
| | Channel: finding a more effective way to reach the customers than the earlier one. | 1 |
| | [a]Zoom-in: Focussing on one specific market sector rather than the whole market. | 1 |
| Others | [a]Complete: Significant change in product, market and financial dimensions but the entrepreneurial team remains the same. | 11 |
| | [a]Side Project: A different business idea parallel and unrelated to the main project becomes the main project. | 2 |
| Total | | 55 |

[a] new pivot type identified in this study

In the case of Pinterest, multiple triggering factors were identified. It used to be a mobile shopping app called Tote, allowed people to browse and shop from their favourite retailers, and also sent them updates when their favourite items were available and/or on sale. The idea of mobile shopping was ahead of its time in 2009, due to the fact that mobile payment solution was not sophisticated at the time. Meanwhile, the founders also discovered that people were more interested in sharing their favourite items lists with their friends and relatives rather than doing window shopping at their favourite stores. The startup company apprehended an opportunity through this unanticipated user behaviour and pivoted towards what we know today as Pinterest.

Table 5 Product related pivots and triggering factors

| Pivot type | Startup name | Triggering factor |
|---|---|---|
| Zoom-in | Flickr | User appreciation of one particular feature of the product |
| | Slack | |
| | Voylla | |
| | BranchOut | Negative customer reaction |
| | Shopify | Identification of a bigger customer need through solving an internal problem |
| | Pinterest | Unanticipated use of product by users |
| | | Wrong timing |
| Technology | Wix | Technology challenge |
| | Android | Targetted market narrowing |
| Platform | appMobi | Identification of a bigger customer need through solving an internal problem |
| | StyleZen | |
| | SymphonyCommerce | Technology challenge |
| Zoom-out | Elto | Unscalable business |
| | Carbon | Flawed business model |

Table 6 Market related pivots and triggering factors

| Pivot type | Startup name | Triggering factor |
|---|---|---|
| Customer need | Jammber | Negative customer reaction |
| | Docker | |
| | Citivox | |
| | ChartBeat | Negative customer reaction |
| | | Unanticipated use of product by users |
| | YouTube | Negative custome reaction |
| | | Wrong timing |
| | Signpost | Unable to compete with competitor |
| | Tagged | |
| | Yelp | Unanticipated use of product by users |
| | Hopper | |
| | BraveNewTalent | Unscalable business |
| | Vidyard | |
| | NowInStore | Influence of investor/mentor/partner |
| | NextBigSound | Flawed business model |
| Customer segment | Eden | Flawed business model |
| | Socrata | Positive response from an unforeseen customer segment |
| | Zealyst | |
| | Groupize | Unable to compete with competitor |
| | Keas | Negative customer reaction |
| Channel | Site59 | Negative customer reaction |
| | | Influence of investor/mentor/partner |
| Zoom-in | Ignighter | Positive response from an unforeseen customer segment |

Table 7 Other pivot types and triggering factors

| Pivot type | Startup name | Triggering factor |
|---|---|---|
| Complete | Twitter | Unable to compete with competitor |
| | Streamline | Influence of investor/mentor/partner |
| | Seesmic | Negative customer reaction |
| | Turntable.fm | |
| | Nextdoor | |
| | Curios.me | Running into legal issue |
| | Fab | Negative customer reaction |
| | | Flawed business model |
| | Woot | Identification of a bigger customer need through |
| Side project | Groupon | negative customer reactoin |
| | | Flawed business model |
| | Life On Air | Side project more successful than main project |

Table 8 Software startups with multiple pivot instances, and triggering factors

| Startup name | Pivot instance | Triggering factor |
|---|---|---|
| Instagram | Product zoom-in | User appreciation of one particular feature of the product |
| | Technology | Technology challenge |
| Handmake Me | Customer need | Negative customer reactoin |
| | Customer segment | Flawed business model |
| MishGuru | Complete | Unscalable business |
| | Customer need | Identification of a bigger customer need through solving an internal problem |
| Paypal Workible | Technology | Wrong timing |
| | Customer need | Influence of investor/mentor |
| RenentionScience | Technology | Technology challenge |
| | Customer need | |
| | Complete | Negative customer reaction |
| | | Unscalable business |
| | Complete | Unable to compete with competitors |

### 4.4.2 Technology Pivot

Technology challenge is a triggering factor to technology pivot. In the case of Wix, it started as a Flash-based website builder when Flash was the best option available for website development before 2011. With the advent of smartphones, mobile devices and introduction of HTML5, Flash was not anymore a viable option for their business because of its performance problem with the smartphones. Due to this reason, Wix pivoted towards providing the website development platform using HTML5.

Technology challenge is also behind the pivot of Android, presented as the emergence of new technology. The original idea behind Android was to provide an operating system for smart cameras that were linked to Personal Computers (PC), and to provide cloud storage to store photos. However during that time, the smartphone and mobile devices industry witnessed high growth, which led to the market of smart cameras shrinking. The camera market became too small for Android business. The combination of the new emergent smartphone technology and narrowing camera market triggered Android to pivot from an operating system for cameras to provide mobile platform (operating system) focusing on handsets.

### 4.4.3 Platform Pivot

Platform pivot can be bi-directional by definition, either from a particular product to an underlying platform or vice versa. However in our case sample, the three startups, appMobi, StyleZen and SymphonyCommerce all pivoted from product to platform. For example, appMobi, originally called Flycast, pivoted from a mobile app for iPhone, Blackberry and Android to a set of tools that support cross-platform development of mobile apps.

The drive to the pivots of appMobi and StyleZen came from solving their internal pain point first. Launched in 2011, StyleZen was an online shopping site. Since the company was struggling with growing its user base without spending a huge amount of money, they decided to use Pinterest to market about their business in order to increase traffic and followers. The team experimented with different types of pinning contents, and identified that there was a solution to increase traffic for their business using Pinterest. The team implemented the solution as a set of tools, and realized that their solution has more power to grow rather than the online shopping site itself, as the co-founder described:

> "Once we figured out the technology, we had a very strong belief that it would be more valuable if leveraged across multiple brands as opposed to my one startup brand. An internal tool turned out to be more valuable than the shell." [StyleZen]

Therefore, the main product of StyleZen pivoted to a Pinterest-optimization platform called Ahalogy for other companies to use Pinterest as a marketing vehicle for their business.

In the case of SymphonyCommerce, technology challenges are the factor behind its pivot. The company initially was a social shopping startup, selling different goods (home, style, living etc.) and requiring users to login through their Facebook account to access the website. Its growth relied heavily on Facebook's Open Graph. However, when Facebook changed Open Graph implementation to make the connected apps less spammy, the viral effect and the large audiences that some apps were enjoying, including SymphonyCommerce, were seriously compromised. Due to this technological change from Facebook, the startup pivoted towards providing a Symphony platform for e-commerce.

### 4.4.4 Zoom-Out Pivot

Two startups, Elto and Carbon, have pivoted in a product zoom-out manner. Tweaky (now Elto) was launched as a marketplace for developers, providing a quick medium for small businesses to make changes in their website with a certain fee. The founder discovered that providing only small changes to the websites was not a scalable idea. This triggered the company to decide to go broader by adding more functionalities and providing more services to their marketplace, and to rebrand the startup's name to Elto ("every little thing online"). The co-founder commented:

> "Tens of thousands of customers later, we realized small tweaks to websites wasn't the way we could add the most value to small businesses, who really want help to grow. Part of that will always be adding functionality, but we're now also focusing on connecting them to marketers and growth strategists." [Elto]

Carbon's zoom-out pivot was due to the flaw in their original business model. It initially started as an on-demand parking that rented space in garages and dispatched valet to pick up cars. However, in order to run the business, they needed intensive investments from the ventures, as the founder described:

> "For on-demand parking, every time you're parking a car, you're spending two, three, even five times more. It's very venture capital subsidized. Carbon didn't want to play that game anymore." [Carbon]

The founder realized that they should provide more services to earn profits. Hence, they pivoted by adding several other services (e.g. laundry pickup) in addition to on-demand valet parking, and became a delivery concierge service for the cars.

### 4.5 Market Related Pivots

Table 6 shows a list of software startups that each contains one pivot instance only and the instance can be classified under the market related pivots. The triggering factors that caused these startups to pivot are also shown in the table.

#### 4.5.1 Customer Need Pivot

The first type of market related pivot is customer need pivot. Negative customer reaction is the triggering factor for several startups, including Jammber, Docker, Citivox, ChartBeat and Youtube. Take the example of Jammber. The company initially started to provide a social app where musicians could find different artists, stylists, photographs etc. Its goal was to become LinkedIn for the music industry. However, while interacting closely with Nashville artists, producers and the musician's union, the co-founders discovered that the musicians spent several hours on paper works to fill different documents etc. They realized this was a real pain point for musicians. A decision needed to be made, as the founder commented:

> BWe had to decide ourselves if we wanted to go the sexy route, or if we wanted to go the money route… We decided to go the money route and do what's best for Jammber. And we're now meeting investors who get that.^ [Jammber]

As a result, Jammber pivoted towards a payment processing and document filing platform between artists and their labels to communicate with each other, share documents and process payments.

For ChartBeat and YouTube, apart from negative customer reaction, their pivots were also driven by other factors. ChartBeat, initially called firef.ly, provided a way for website owners to see what users were browsing around their websites, and let the same webpage users chat with each other. However, instead of browsing the website, the users started chatting with each other and discussing about firef.ly most of the time. This unanticipated use of the system, together with low traction of the intended product, triggered the startup to pivot to ChartBeat which provides different user information (how many people were on their site, where they were coming from and what they were reading) to the website owners. In the case of YouTube, it initially started as a video dating platform to find possible dates via videos. The idea was ahead of its time in 2005. Moreover, the idea did not get much user attraction. Hence, it pivoted towards a platform to share videos online.

The impact of competition can be crucial, as shown in the cases of Signpost and Tagged. Both had to pivot the problem they initially wanted to solve for their customers. In the case of Signpost, the startup originally was a website for daily deals, similar to Groupon. However Groupon was a stronger player in this field, which forced the founder of Signpost to search for new direction. Through talking with different local businesses, Signpost discovered that, although there were many local sites available e.g. Yelp, Yahoo local, and Google local, there was no tool available to update the business profiles on these sites. The founder realized this was an unsatisfied need and pivoted towards helping small to medium sized businesses to better market themselves online.

Yelp and Hopper discovered real customer needs due to the unanticipated use of their products by users. For example, Yelp intended their emailing system to be used by users to connect with others for recommendation about local businesses. However, the users used the system to write reviews about the local business. This emergent new mode of using the system soon caught the attention of the founders, through which they identified a different yet more promising need to be catered, and pivoted towards providing reviews about local businesses.

The realization that their businesses were unscalable pushed BraveNewTalent and Vidyard to seek for new problems to solve. The initial idea behind BraveNewTalent was based on the assumption that people wanted to follow the companies for whom they would like to work in the future, and companies wanted to educate potential candidates on how they should work. It intended to be a social recruiting platform. However, the startup did not figure out how to scale their business idea as the founder described:

> "We realized we were trying to build communities around recruitment content. But trying to build user engagement around transactional content like that doesn't work. Job seekers just wanted jobs, and recruiters just wanted to fill positions. So the model didn't scale just focusing on jobs." [BraveNewTalent]

Consequently, BraveNewTalent pivoted from a social recruitment platform to social learning platform and primarily focusing on enterprises settings.

The story of NowInStore presents an example of customer need pivot triggered by yet another factor. Initially, the startup developed an app that allowed different brands to showcase their products in a portfolio. The startup was accepted in a New York based technology accelerator. During their stay in New York, the founders had meetings with dozens of investors. Due to the influence of investors, they realized what could be more effective, and pivoted to a platform that helped businesses create marketing content by leveraging the data from online stores. The founder commented:

> "The product has neatly positioned itself as a marketing platform for small and medium-sized businesses (SMBs), leveraging smart data, and it's going really well now." [NowInStore]

In the case of NextBigSound, it was flawed in their original business model, which triggered the pivot. The original idea behind NextBigSound was to develop a website for music passionate people, where they could create fantasy record labels to Bsign^ artists. Although they had thousands of users, they could not generate enough revenue from this user base to survive. To find the real need of the musicians and artists, the founding team remained the focus on the music industry, and tried to identify the mantra behind popularity of bands from garage to a big hit. The new idea was to track social media to measure popularity of new and up-coming music. Hence, they pivoted, and became a company which provided data analytics to artists, producers and labels.

### 4.5.2 Customer Segment Pivot

Five startups, Eden, Groupize, Keas, Socrata and Zealyst, are classified under this category, and each pivot is triggered by a different factor except Socrata and Zealyst, who have the same triggering factor. Eden initially started with providing technical support to consumers to solve their information technology related issues – similar to geek squad. The startup was mainly focused on providing services to end consumers to generate revenue. However, they soon realized that their revenue model was not working as they expected, because the consumers were more sensitive to price than they expected. Meanwhile the founder also discovered that their main revenue was generated through the business customers (enterprises), so they pivoted customer segment from consumers to business. The CEO commented on this pivot:

> "It's tempting as a founder to have a consumer business idea, but once we noticed the best part of our business was in B2B, we had to pivot. Serving businesses tends to have higher frequency and higher profit." [Eden]

Groupize originally targeted at consumer business by generating demands through creating white-label agreements with multiple travel agencies. Unable to compete with their competitors in the travel management service areas, they shifted their focus towards providing a group-meeting solution for hotels, meeting planners and travel management companies, and created partnerships with different hotel chains and independents. In contrast, the pivot of Keas was due to negative customer reaction. Initially planning to provide personalized health care places for consumers by leveraging their personal health data, Keas ended up providing workplace wellness program for different businesses since their idea did not get much traction in consumer market.

Socrata and Zealyst pivoted due to the same triggering factor. For instance, Socrata was launched to create a cloud-based database for small to medium sized businesses. The original idea was to put one's database in a cloud and let someone else manage it. During the presidential campaign in 2008, the president's campaign team used this platform to put contribution data online. This positive response from previously unforeseen customer segment made the founder recognize that government can put data online by using a cost effective solution like cloud computing. The startup pivoted accordingly towards a cloud based offering for open data government.

### 4.5.3 Channel Pivot

Site59 presents an example of channel pivot. The initial idea of Site59 was to create mini-vacation packages by combining last minute offers from different air travel, hotel accommodation and other travel related services. However, the idea did not go viral as expected, and the customer acquisition rate was low. One of the investors suggested Site59 to change the distribution channel to reach customers, using the business to business to consumers (B2B2C) model. Following the suggestion of their investor, Site59 pivoted the channel to reach their customers and their new service was to prepare last minute vacation packages for different airlines and travel portals which eventually reached the end customers they intended to serve initially.

### 4.5.4 Zoom-In Pivot

Market zoom-in is a new pivot type emerged from our study. It is a type of pivot where a startup narrows down its target market from a broader one to a more specific market segment. Ignighter is an interesting case of such a pivot. The primary aim of Ignighter was to develop a dating website for users. The targeted audience is general, however the founders expected to receive positive response from the US, their home market. Unexpectedly, the idea got promising attraction from customers in Asian markets, especially in India. The founders carefully analysed the demographic data, and identified the promising user growth in India as compared to any other country. As a result the founders decided to focus on this market segment and made the pivot, stating that Bwe are an official Indian dating site.^ [Ignighter]

### 4.6 Other Types of Pivots

Table 7 shows a list of software startups that each contains one pivot instance only and the instance cannot be classified readily under either product or market category. The triggering factors that caused these startups to pivot are also shown in the table.

#### 4.6.1 Complete Pivot

A complete pivot is a pivot where an entrepreneurial team has to come up with a new innovative idea after their initial innovative product/idea was outplayed due to different factors (by their competitors) e.g. big companies started working on that idea and attracted their niche markets. This pivot implies significant change in one or more aspects of a startup, including product, targeted market and finance. The only unchanging element is the entrepreneurial team that carries on the learning from the past experience to the new directions.

Twitter is an example of complete pivot. It initially started as a podcast service (Odeo) to allow sharing and recording of podcasts. Then Apple iTunes started to fill this gap, leaving behind the Odeo service. As they were unable to compete with Apple iTunes, the startup team had to brainstorm to find a new direction, and came up with a new messaging service called Twitter.

The complete pivot of Streamline can be attributed to their mentors while they were in a Techstars Seattle program. Initially the startup aspired to improve and streamline the customer service industry by reducing the waiting time during call-ins to customer service centres. After they were selected into the Techstars Seattle class, with the advice and guidance of Techstars mentors, they completely changed their initial idea, and pivoted towards helping brick and mortar owners in their decision making related to expansion. The founder described the new idea as "helping retailers expand intelligently to the right physical locations, because one mistake on a location could destroy a small retail concept. We're solving this pain." [Streamline]

Three complete pivot cases, Seesmic, Turntable.fm and Nextdoor, were triggered by negative customer reaction. Take Seesmic as an example. It started to be a video-based Twitter. The founder realized that people would prefer to tweet in words rather than contributing in a discussion by recording a video message, which is too much of a hassle for engaging in a discussion. Based on the negative customer reaction (no traction, difficulties in recording coherent video message), they changed their direction completely, and pivoted towards developing a social media client application particularly focusing on enterprise services.

Curios.me had to pivot completely because it ran into legal issues. Originally, it was a Facebook fan-site network known as CrowdRally. The Facebook lawyer's team identified certain uses of CrowdRally as inappropriate. Curios.me pivoted towards providing a question-answer site for college students due to the little chance of resolving the legal trouble.

Fab.com pivoted completely due to two factors. Starting as a social network targeting the gay community, the startup did not get much traction, was stuck at one point, and unable to reach the revenue point that they had projected. The other jobs the founders had in the design field hinted them when they searched for new directions. Consequently Fab.com pivoted towards selling handpicked home goods, clothing and accessories. This new direction took off and it has now become the well-known model fashion website.

Solving an internal problem sometimes can lead to a complete pivot. In the case of Woot, it started as a wholesale electronic distributor and wanted to clear out their unsold inventory. While solving own inventory problem, Woot discovered a unique model for online shopping website combining both the need of urgency and scarcity, and pivoted towards software industry.

### 4.6.2 Side Project Pivot

Side project is a special kind of project that runs parallel to the main project of a software startup, but may be based on a different even unrelated business idea and target at a different set of customers. Groupon is a well-known example of side project pivot where side project outshines the main one. However the deeper reasons were because of the issues with the main project. Groupon initially started as 'The Point': social campaigns to collect fund for good causes. Campaigns were only successful when a certain tipping point was reached. However this project did not get much user traction, and there was no clear revenue model therefore it was difficult to monetize the idea. However, the side project the team started in parallel, using the same tipping point but for group buying and local deals, attracted more users. Eventually the side project took off and it has now become the daily deal website famously known as Groupon.

Life On Air is also an example of pivot where side project overshadows the other projects and becomes the main project. The side project was called Meerkat, an app enabling users to stream live video to their Twitter followers with a single click. It is different from the main business of the company which was another app called Air. Meerkat's usage shot up almost immediately after launch and topped the list on Product Hunt, a site that allows technology enthusiasts to surface and vote on new tech. Over 3 days roughly 15,000 people used the Meerkat service. As a result the founder of Life On Air decided to focus the team on Meerkat fully. He commented:

> "Meerkat is the embodiment of the ability to run really fast, look up and see whether you are going the right way, and if not redirect yourself." [Life On Air]

## 4.7 Multiple Pivots

We identified six startups that evidenced multiple pivots. Table 8 shows a list of software startups that each contains two pivot instances together with triggering factors.

Instagram provides an illustrating example of a startup which evidenced two product related pivots, zoom-in and technology, during their journey. Instagram originally was a location-based service called Burbn, combining features of Foursqaure (photo share app) and Mafiawars (game). Users could earn points for hanging out with their friends, and share pictures inside of the app. However the users appreciated only one particular feature of the product, photo sharing. The founders also realized the need of focusing on this specific aspect (a product zoom-in pivot), as the co-founder explained this situation:

> "I've heard that Plan A is never the product entrepreneurs actually end up with. I didn't believe it. In many ways, Burbn was getting a bunch of press, but it wasn't taking off the way we thought it would. We found people loved posting pictures, and that photos were the thing that stuck. Mike, my cofounder, and I sat down and thought about the one thing that made the product unique and interesting, and photos kept coming up". [Instagram:2]

The service was initially a browser based mobile app developed in HTML5. However due to latency issues in HTML5, they pivoted towards providing an iOS only app for iPhones. The main reason behind this technology pivot was the limitation of the existing technologies.

Handmake Me is a special case which evidenced two market related pivots: customer segment and customer need. Their original idea was to provide a portfolio service for professional creatives, called Host My Portfolio. However, the idea did not bring any revenue. They realized that most of the products on their service were craft-related. They pivoted towards hobbyists (customer segment) and created a Breverse marketplace for handmade gifts^ called Handmake Me. It allowed anybody who wanted to buy an authentic handmade gift to request what they wanted and how much they would pay, and craft-makers would then bid for the task. However, nobody was requesting anything. In order to solve this, they changed their direction towards a more conventional marketplace. They implemented various customized option for buyers, and implemented a strict quality assurance criteria. BIt worked. People are buying, people are selling and we are revenue generating at last.^ [Handmake Me]

Another example of multiple pivots is MishGuru, which pivoted twice: complete and customer need. The original concept was to develop a platform that allowed users to custom design and print their own 3D horse shoes. During their work in a lab accelerator program in New Zealand, they discovered that their idea was not scalable because their targeted market (horse owners) was not conducive for rapid growth. Hence, they pivoted in a complete new direction. The founder explained the new idea:

"We started playing around with an idea for collaborative video making in between friends. As part of that, we spent a couple of days building a really basic MVP using Snapchat as the base platform to build it off since everyone already had the app." [MishGuru]

While working on the idea of collaboration video making, they stumbled upon another challenge. They were unable to find a solution to manage content, and drive user engagement on snapchat. They developed their own solution to manage these kinds of activities. Now, they have become a content management system for Snapchat, where users can create storyboards, build campaigns etc., and publish on Snapchat.

PayPal and Workible are examples, where both evidenced technology and customer need pivots. Although they evidenced similar pivots, the triggering factors were different. In the case of PayPal, timing issue was behind the pivots. PayPal pivoted from a money exchange solution for Palm Pilot users to online payment, since mobile payment was Bstill'the future' 12 years later^ [Paypal:2]. The new customer need that PayPal pivoted to, online payment, was influenced by the merge with X.com and partnership with eBay. In the case of Workible, it initially started as a job site, which helped job seekers to find the job based on their availability. The solution was a desktop based website initially. However, as the founders commented:

"We had what we call now a blinding flash of the obvious, but it was a bit of an epiphany that if we were going to build a world class tech business we could not do it with a website we built out of India. And overwhelmingly we got this message that everything was going mobile and everything had some social component." [Workible]

At the same time, the founders did a massive amount of research into the market and started asking people what their biggest problems were. They learnt that the main issue was the difficulty to find quality people fast. Mobile is the perfect solution for that because it allows customers to connect with people instantly. After realizing the real need of the customers, the founders pivoted towards providing a mobile platform that allowed companies to instantly find the candidates that matched a set of criteria.

RetentionScience is an interesting case which evidenced two complete pivots before founding RetentionScience. Initially it provided independent artists a platform where they could promote niche brands and products via social media. Although the founders contacted different channels (working with YouTube celebrities, sponsoring local concerts, etc.), their business proved to be unscalable. They also discovered that the customers were reluctant to appear as sell-out by promoting different brands. The unsalable business and negative customer reaction triggered their first complete pivot. They pivoted towards providing a social media-based analytics and referral platform for e-commerce businesses. The second complete pivot happened due to the competitors. The founder discovered that there were many well-funded startups working in the same area, and there were little chance that they could acquired the funds to compete. Without funding, they could not accelerate their product development and increase user growth, hence unable to compete with their competitors. Therefore, they pivoted completely again towards a retention automation platform that used artificial intelligence techniques, to engage existing customers and increase customer retention.

## 5 Discussion

When the conditions are uncertain and chaotic, failure is almost inevitable (McGrath 2011). It is rightly applicable in the context of software startups that work under the conditions of extreme uncertainty. A common element behind all the pivot cases we studied is validated learning. Without it, the pivot decisions would be ungrounded (Ries 2011). Startups obtain validated learning through failures, therefore making Bintelligent failures^ (McGrath 2011). As shown in the cases reported in Section 4, the studied software startups adopted this validated learning process and used the acquired knowledge to set right directions. That is why in the context of startups, failure is viewed positively, and failing fast and failing often is the mantra of most lean startups. If startups do not learn from their failures, there is a high probability that they would eventually fail permanently (Richardson 2011).

### 5.1 Reflection on Pivot Types

There are few studies related to pivots in software startups as described in Section 2. Our study extends the scope investigated in Van der Van and Bosch (2013) and Hirvikoski (2014), and adds new knowledge regarding pivots, pivot types and triggering factors. The study by Terho et al. (2015) presents different types of pivots as described by Ries (2011), and how they affect the business model. Our findings extend this study by providing further types of pivots (e.g. complete pivot etc.), and also provide the categorization of different factors that trigger pivots in software startups.

Among the pivots identified in our study, the most common is customer need pivot (17 out of the 55 pivot instances). This is not surprising in the sense that it is consistent with the nature of startups in general and software startups in particular. While working with highly dynamic and uncertain technologies and building innovative products, software startups are striving to find the real and unique customer problems that are worth solving. In order to better understand customer needs and identify real problems, software startups need to pivot relentlessly.

The significance of customer need pivot reveals the importance of identifying the right problem, first. In order to develop something valuable for their customers, startups need to understand their problems (Blank 2005). However, the existing studies show that software startups tend to ignore the identification of right customer problems, instead focus on developing solutions and investing in product/market fit prematurely (Giardino et al. 2014, 2015). It is highly probable that the initial assumptions towards customers prove to be wrong. This is one of the reasons of customer need pivot, where startups should pivot because the problems that they identified are not real pains for customers. This is manifested in the case of Yelp, Hopper and Jammber etc. where they pivoted subsequently according to different customer needs they discovered.

In the course of better understanding the market, software startups often discover that, even though the problem they want to solve is real but it is not the problem of the customer segment they have initially presumed. 6 pivot instances made the customer segment pivot, the second most common market related pivot type that our study has identified. Startups do not know their potential customers in advance, and risk spending too many resources to come up with a product that fails to achieve product/market fit. However, every failure has some lessons to be learnt. This learning can be helpful to identify new targeting customers and then pivot towards them, as manifested in the cases of Eden, Socrata, Groupize, etc.

Product related pivots are also important pivot types. 17 pivot instances in our sample are related to product. Software startups have to reconsider their products and different features in order to find the problem/solution fit and/or product/market fit. This often leads towards product related pivots. Among different product related pivots, the zoom-in pivot is relatively more common than other product related pivot types. It often happens that customers are more interested in a particular feature rather than the whole product. Pinterest and Flickr are good examples of product zoom-in pivot. Ideally, instead of wasting resources and building a complex product with lots of features, it is better to focus on one feature that actually gained the attractions of the customers and build it first. However it is not easy to understand which can be the valuable feature to build first. Ries (2011) suggests to develop MVP in order to test the hypotheses related to product, business model, and/or engine of growth. It may be beneficial for entrepreneurs to use MVP to decide the candidate features to include in their product offerings. By building MVPs, entrepreneurs have an initial set of features that are appreciated by the initial users.

The opposite of product zoom-in pivot is product zoom-out pivot, which reflects the need of achieving the problem/solution fit. It is possible that software startups have identified the right set of problems, but their products are still incomplete. They need to expand their solutions to add more features. Elto did zoom-out pivot by providing additional functionalities to its original product, and secured a better place in the market.

Another important product related pivot type is technology pivot, second most common within the product dimension, which reflects the role technology plays in software startups. Software startups are prone to technology pivots due to the fact that they are building technology intensive products. Often technology pivots are driven by the need of software startups to be always at the cutting edge of technological advancement. This is manifested in the cases of Wix, Workible, Android and Instagram. Sometimes existing technology has some performance issues, and technology pivots help startups to come up with an improved solution, as shown in the case of Instagram.

In order to support their products and make solutions complete, software startups often develop both products and supporting platforms. The platforms support their core products. Sometimes it happens that the platforms solve larger problems than their original products do. Therefore platform pivots are desired. StyleZen is one such example. The startup pivoted towards a platform to support online shopping businesses instead of becoming yet another shopping website. It is worth mentioning that platform pivot can be also from a platform to a product running on the platform. However, we could not find evidence in our case collection. One possible explanation can be that this direction is not as frequent as the product to platform direction.

In terms of the scope of change and the amount of effort and resource needed, no pivot type is more demanding than complete pivot. This is a new pivot type identified in our study. It is the second most common among all pivot types (11 out of 55 pivot instances). We term it complete pivot since it is related to almost all the aspects of a startup, including product, market and financial, with only the original team as the rooting element in the pivot, which ensures the learning from previous failing experience is maintained. Famous companies such as Twitter, Fab.com, and Turntable.fm all went through significant changes in their business before they found successful and sustainable business model to scale.

Side project pivot is another interesting new pivot type. Even though working under a high-pressurized and extreme chaotic environment, many software startups run one or more side projects simultaneously that are generally not related to their main ideas. A side project is a project that runs parallel to the main project, but may target at a different set of customers. These side projects may become main projects when outshining them. Groupon is a good example that was initially started as a side project. Therefore it is arguably beneficial to have a side project parallel to the main product development project. Further studies need to be conducted in order to explore the importance and implication of side project, and the cost associated to running such parallel project, especially in the software startup context.

The third new type is market zoom-in pivot, which is demonstrated by the Ignighter case. It is a reflection of striking the product/market fit. It is often suggested that, to start with, a startup should find its focus and niche market, identify the early adopters of their product. This type of pivot shows the need to do so. However, since we only found one instance in our case collection, the robustness of this new pivot type is yet to be tested.

Another finding worth mentioning is multiple pivots which can happen either simultaneously or separately. Some pivots may be closely linked and there is a possibility that chain reaction occurs, which means one pivot triggers several other pivots, known as Bthe domino effect^ (Terho et al. 2015). This chain reaction is manifested in the case of Instagram (product zoom-in and technology pivot) and Workible (customer need and technology pivot), where two pivots occurred simultaneously. One startup may have several pivots spread across their courses of development too, which is revealed in the case of RetentionSciene (two separate complete pivots). This indicates the importance of constantly checking and correcting the directions until a startup obtains a sustainable business model.

A pivot can also be related to business architecture, value capture, and engine of growth or social aspects. However our study does not find any case that evidences these kinds of pivots. One reason could be that these pivot types may be less common in comparison to other types, therefore not appearing in our case collection. Why these are less common types, however, is beyond the speculation. Broaden the scope of cases may surface these pivot types in software startups, verify if they are indeed less common, and unveil why they appear less frequently as pivot types.

A study by Giardino et al. (2015) reveals that building an entrepreneurial team is one of the prominent challenges for software startups. Our results show that, however, there is a knowledge gap regarding the pivots related to the team dimension. The same is true for the financial dimension. Even though several pivot types are suggested by Ries (2011), our results did not yield any related cases. Both team and financial are important dimensions of successful software ventures (Macmillan et al. 1987), and further investigation is needed in order to gain any insights in these aspects.

### 5.2 Reflection on Triggering Factors

The majority of the triggering factors listed in Table 3 are considered external factors, which are events occurring beyond the control of a startup. This implies that for many of the studied startups, the major pivots they made were more reaction to what happened externally rather than purposefully design change as suggested by Ries' definition of pivot (Ries 2011).

Table 3 shows that negative customer reaction is the most common factor triggering pivots. Slow user acquisition, low user retention rate and no growth are some manifestations of negative customer reaction, and show that startups are unable to achieve the product/market fit. Negative customer reaction works as a first litmus test for the startups to decide whether they are solving the right problem for the right set of customers or not, and consequently whether they should pivot or not. In the case of Seesmic and Jammber, both startups reacted to negative customer reaction and pivoted subsequently.

In order to come up with innovative and cutting-edge products, software startups have to compete with other competitors and especially with big companies. It is the second most common external factor that triggers software startups to pivot. The big companies have much more resources than what software startups can wield. They can implement innovative products rather quickly as compared to startups. Twitter stumbled upon this challenge when their initial idea of offering podcast services was outplayed by Apple with the launch of iTunes. They pivoted drastically.

Thriving in technology uncertainty is the top challenge faced by early stage software startups (Giardino et al. 2015). Accordingly, technology challenge is a common factor that triggers pivot, and the pivot types are generally related to technology or product, such as shown by the cases of SymphonyCommerce and Wix.

One of the contextual features of software startups is that they are heavily influenced by stakeholders and investors (Paternoster et al. 2014). The suggestions from the investors/mentors/partners greatly affect the development processes of software startups, and may eventually change their course. It is probable that a software startup has a good technology idea, but their investors, mentors or partners have a different vision, which affects the overall direction of the startup. Investors/mentors may have valuable suggestions that are worth listening. NowInStore and Streamline are good examples of opening up to the suggestions from investors/mentors, and changing their directions accordingly.

Meanwhile, our findings also reveal that a flawed business model is a prominent internal factor that triggers software startups to pivot. Scaling a business that has flaws in their business model, or in other words, "pre-matured scaling" (Giardino et al. 2014), may lead to the eventual failure of a startup. Software startups may avoid failure by identifying the flaws in their business model earlier and pivot accordingly. Low or no revenue, or high acquisition cost indicates a flawed business model and consequently the need to pivot, as demonstrated in the cases of Groupon and Fab.com.

As shown in Tables 5, 6, 7 and 8, there is no linear and one-to-one relationship between pivot types and factors triggering pivots. Different factors trigger different types of pivots in software startups. For example, unanticipated use of product by users is a factor that triggers Pinterest to product zoom-in pivot, while in the case of Yelp, the same factor causes it to pivot to different customer need. Similarly, it is probable that multiple factors collectively trigger a pivot in a software startup. Instagram, Fab.com, and Groupon are prominent examples of startups where pivot occurred due to several factors. It needs to be emphasized that the list of triggering factors in Table 3 is not exhaustive. There may be other triggering factors yet to be discovered.

Last but not least, our findings on the major types of pivot and triggering factors causing pivots allow us to reflect upon the role of the unique nature of software product plays in software startup pivoting. For example, due to the flexible and modifiable nature of a software product, product zoom-in and zoom-out pivots should be relatively easy to implement for software startups than for startups that produce physical products, such as hardware or medical devices. Similarly, since it is common that the use of software is always appropriated by users, which is generally beyond its designed use (Mendoza et al. 2010), software startups should be more watchful to the triggering factors such as unanticipated use of product by users.

## 5.3 Validity Threats and Mitigation

### 5.3.1 Threats to Construct Validity

This aspect of validity threat refers to what extent the operational measure really represent what is investigated related to the research questions (Runeson and Höst 2009).

One threat to construct validity stems from the use of Bpivot^ only in the search string to find pivot cases. There may be cases that represent pivot but do not use this specific term. They may rather use general terms such as "changing direction", "changing strategy" or "strategic change". As a result, we may have missed some important pivot cases. However, pivot has become a popular and common term used in startup communities, together with the Lean Startup movement. In addition, changing business direction and strategy has a broad meaning and does not reflect the specific implication of pivot which emphasizes on the change that allows Bto test a new fundamental hypothesis about the product, strategy, and engine of growth^ (Ries 2011). Therefore, the probability of using these broad terms to get relevant cases is low.

One limitation of the study is the exclusion of known pivot cases because of lack of information about their pivoting story e.g. Facebook and Nokia are two world famous companies, but we did not get sufficient information to find the triggering factors for their pivots.

### 5.3.2 Threats to Internal Validity

Internal validity threat refers to the broader problems of making inference (Yin 2003).

One internal validity threat to our findings is also related to the secondary data we employed since we have no control over the quality and accuracy of the data, which leads to a weak basis for data analysis. In some cases the pivot instances were described in the interviews of the founders, in other cases were recounts from the writers of news articles who did not have direct working experience in the startups being reported. This validity threat was partially mitigated by data triangulation, using multiple sources of the same cases wherever possible. The benefits allowed by secondary data, e.g., obtaining a large number of real world pivot cases (49 software startups representing 55 pivot instances) with relatively less effort in a short timeframe, out-weighted this potential validity threat.

Another threat is the quality of the selected cases. To reduce the personal bias in case selection, we defined a review protocol with quality assessment criteria (as described in Section 3). This review protocol was discussed and agreed among the research team to cover maximum studies. Researcher triangulation was applied to increase the internal validity. Two researchers conducted the key review tasks independently. As for the selected cases, not all of them were equally informative. There were few pivot cases which provided more details (e.g., RetentionScience) as compared to others (e.g., YouTube). However we ensured that all the key information about each pivot case was available.

To ensure the replicability of our study, we provided as many details as possible and necessary in Section 3. The data collection and analysis steps reported are replicable. However, due to the dynamic nature of the Web and Google search engine, the search results using the same search string reported in Section 3 may not be replicated completely with the lapse of time. To mitigate this potential threat, we published the cases we obtained from the search results in the online repository (the link is given in Section 3 Step 6).

#### 5.3.3 Threats to External Validity

This aspect of validity refers to what extent it is possible to generalize the findings (Runeson and Höst 2009).

The first potential threat to external validity is posed by Google search engine on the accessibility of search results. In our case we had access to 783 links, which were the top-ranked query results according to its search algorithm. We could not know if these 783 links were representative of the total search results (1,070,000).

A pivot action taken by a startup is not guaranteed to be successful and lead to eventual successful startups. However, our sample of pivot cases may be biased towards successful pivots and startups due to the fact that these pivots were published online, many times as the examples of successful pivot. This bias may pose a threat to the generalizability of our findings to the general population of software startups.

Another potential threat to generalization is the representativeness of the resulting collection of cases. The 49 software startups come predominantly from the United States (37 out of total 49 software startups). Since the US is arguably the world leader in innovation and technology as represented by Silicon Valley and other startup ecosystems, it can be argued that our sample is representative to certain extent. As far as pivot is concerned, we believe that the demographic distribution should not impact on the ways startup directions are changed.

The limited number of cases on different types of pivot did not allow us to conduct any quantitative analysis, such as meta-analysis, to identify the relationship between consequence and antecedence of pivots. Besides, much case context information was reported in a large deviance in term of business stage, company size, product type, industry domain and etc. This presents a potential threat to the generalizability of the pivot types and triggering factors to different contexts. However the main objective of this study is to explore software startup pivot phenomenon qualitatively. We targeted at theoretical generalizability rather than statistical generalizability.

One limitation of the study is regarding the generalizability of the identified pivot types and triggering factors. The 49 software startups in the sample do not cover all the business domains (e.g. missing real time application, safety critical application etc.). Moreover, not all the pivot types and triggering factors are manifested in the equal amount of cases, some only have one or two corresponding cases, such as market zoom-in and side project as pivot types and running into legal issues as a triggering factor. Future work is needed to further explore these pivot types and triggering factors to increase their generalizability.

## 6 Conclusions and Future Work

Software startups are producing innovative software products and solutions in dynamic markets using cutting-edge technologies. To achieve success, they need to continually make the decisions of pivoting. This paper provides an initial classification of pivot types that software startups have made through a case survey study of 49 software startups that pivoted, and the factors that triggered software startups' pivots.

The study results show that customer need pivot is the most common type of pivot among all types, while zoom-in and technology pivots are major product related pivot types. Our findings extend the existing knowledge of pivot by introducing three new pivot types: market zoom-in, complete and side project. A knowledge gap regarding team and finance related pivots is also underlined. Among the multiple external and internal pivot triggering factors identified, negative customer reaction represents the most common reason why software startups pivot, followed by the flawed business model of software startups.

Our findings have implications to both software startup practice and research. The empirical evidence from the analysed cases suggests that software startup teams should gather maximum knowledge and consider failure as an opportunity to obtain validated learning. Considering the chaotic and unpredictable environment of software startups, the validated learning will be crucial to drive business and product decisions in order to proceed in the right direction. Both the identified pivot types and triggering factors can be utilized by startups to make more informed decisions on when and how to pivot. It also implies the use of different metrics to track customer reaction, product usage, etc., to support informed decisions regarding pivoting. To get the actionable metrics is important especially in the case of product zoom-in pivot, where data analytics can help to identify the usage of different features.

As for the implication to research, our study provides a set of empirically validated or derived pivot types that can serve as the conceptual basis for future studies on the pivot phenomenon. Similarly, the identified triggering factors and their definitions can be utilized as the conceptual basis to formulate hypotheses in explanatory studies that explore the linkable between the trigger factors and pivot types.

Our study opens up several new avenues for future research on software startups and pivoting. The first direction is to collect primary data to validate the pivot types and triggering factors identified in this study. Primary studies, especially case studies, can also allow us to collect more contextual information on software startups, such as which markets they targeted at when pivots happened, at what stage of product development, in order to achieve more in-depth understanding of the pivot types and triggering factors. Further studies can also explore the actual process of a pivot and its consequences (both technical and business). Another future direction is to study pivoting in software startups in a longitudinal manner. Our study took the snapshots of pivot instances of the studied startups. Future studies can consider the business and product development life cycles of a startup, and investigate at which stage pivot most probably occurs, is most beneficial, and costs least. Taking a longer time span, future work can also compare how software startups today pivot differently than those, e.g., 40 years ago, and how triggering factors vary over the time. In addition, an entrepreneurial team is one key element of a startup and future studies could be conducted in order to address how team size and structure are related to different types of pivots. From a product perspective, it could be interesting to explore how the characteristics and attributes of software, such as complexity and modifiability, may influence pivoting decision and process. Moreover, these cases may indicate the need to identify and track the actionable metrics, and role of these metrics while making pivoting decision. Last but not least, our findings open up an interesting direction of research regarding relationship between triggering factors and pivots, to investigate if there is any causal-effect patterns between triggering factors and pivots. Further studies need to be conducted to identify other triggering factors causing pivots in software startups, and investigate how same triggering factors cause different types of pivot.